\documentclass[showkeys,showpacs,superscriptaddress,nofootinbib]{revtex4-2}
\usepackage{amsmath}
\usepackage{amssymb}
\usepackage{dutchcal}
\usepackage{graphicx}
\usepackage[normalem]{ulem}
\usepackage{xcolor}
\usepackage{float}
\usepackage{slashed}

\allowdisplaybreaks

\begin{document}

\title{Realistic classical charge from an asymmetric wormhole
}
\author{
Vladimir Dzhunushaliev
}
\email{v.dzhunushaliev@gmail.com}
\affiliation{
Institute of Nuclear Physics, Almaty 050032, Kazakhstan
}
\affiliation{
Department of Theoretical and Nuclear Physics,  Al-Farabi Kazakh National University, Almaty 050040, Kazakhstan
}
\affiliation{Academician J.~Jeenbaev Institute of Physics of the NAS of the Kyrgyz Republic, 265 a, Chui Street, Bishkek 720071, Kyrgyzstan}

\author{Vladimir Folomeev}
\email{vfolomeev@mail.ru}
\affiliation{
Institute of Nuclear Physics, Almaty 050032, Kazakhstan
}
\affiliation{Academician J.~Jeenbaev Institute of Physics of the NAS of the Kyrgyz Republic, 265 a, Chui Street, Bishkek 720071, Kyrgyzstan}

\begin{abstract}
Within Einstein-Dirac-Maxwell theory, we consider a wormhole solution supported by a complex non-phantom spinor field with a bare mass of the order of the Planck mass
(which provides a nontrivial spacetime topology and an intrinsic angular momentum),
an electric field (which provides a charge of the system), and a magnetic field. This solution describes an asymmetric wormhole connecting two different asymptotically flat
spacetimes (two universes) in which there are in general different observed masses and charges.
It is shown that, by suitably adjusting the values of free system parameters, at one end of the wormhole, one can obtain the values of the observed mass and charge typical of the Standard Model particles,
whereas at the other end of the wormhole these physical quantities acquire the Planck values. 
Such a configuration incarnates Wheeler's idea of ``mass without mass'' and 
``charge without charge'', and can be thought of as a model of a classical charge possessing a spin.
\end{abstract}


\date{\today}

\maketitle 

\section{Introduction}

The Standard Model (SM) of particle physics, incorporating the weak, strong, and electromagnetic
interactions, is a quite successful theory that enables one to predict the interactions of quarks and leptons with high accuracy. 
However, it has a number of weaknesses and is believed to be inadequate for describing Nature at a truly fundamental level.
In particular, it cannot describe the origin of the fermion masses, which arise in the SM as a result of the Yukawa interactions
with the Higgs doublet: there is a huge hierarchy of fermion masses (for example, the ratio of the mass of the top quark 
to the mass of the electron is $\sim 10^{5}$), and correspondingly of magnitudes 
of the Yukawa couplings to the Higgs (the fermion mass hierarchy problem). 
Even more striking difference is
in magnitudes of the masses of the electroweak gauge bosons/Higgs and the Planck mass (the so-called gauge hierarchy problem). At the classical level, 
this is clearly evident from the fact that the ratio of the Fermi constant $G_F$ (determining the weak interaction scale)
to the Newtonian constant~$G$ (determining the gravitational force strength) is $G_F/G\sim 10^{33}$.

The hierarchy problem was often addressed in the literature in different aspects by involving new physics, 
including supersymmetry~\cite{Feng:2013pwa}, composite Higgs theories~\cite{Bellazzini:2014yua}, brane models~\cite{Arkani-Hamed:1998jmv,Randall:1999ee}.
Whatever new physics is employed to extend the SM at high energies, a common approach to address 
the hierarchy problem implies the use of some effective field theory.
Such a theory, providing a low-energy description of Nature within the SM, 
can involve only a finite set of parameters coming from unknown UV physics. 
 One of such parameters is the mass of the Higgs boson. If the electroweak and gravitational forces are treated within quantum field theory, 
it is necessary to include quantum gravity loop corrections to the Higgs mass, and these corrections are naively estimated to be of the order of the Planck mass.
To eliminate the divergent UV contributions to the Higgs mass,
in ``natural'' theories new physics is introduced right above the Fermi scale (see, however, Ref.~\cite{Giudice:2013yca}
where this naturalness principle is questioned due to the lack of signatures of new physics at the TeV scale). 
However,  a big difference between the Fermi and the Planck scales remains nevertheless
unexplained.

Consistent with this, here we consider a new mechanism that enables us, on the one hand, to use in the parent Lagrangian the
fundamental Planck mass (as a bare mass) and, on the other hand, to obtain a measured effective mass of a localized field object that is typical of the SM particles.
For this purpose we employ rotating wormhole solutions in general relativity supported by a complex non-phantom spinor field
and electromagnetic fields obtained in our recent paper~\cite{Dzhunushaliev:2025ntr} (see also the pioneering work~\cite{Konoplya:2021hsm} 
devoted to nonrotating wormholes in Einstein-Dirac-Maxwell theory). 
Those solutions are asymmetric, regular, asymptotically flat and carry nonzero intrinsic angular momentum,
enabling us to describe localized gravitating configurations possessing a nontrivial spacetime topology and physical characteristics 
corresponding to the Planck scale.
The physical properties of the resulting configurations are completely determined by the values of three input quantities: 
the throat parameter, the spinor frequency,  and the electromagnetic coupling constant. The wormholes 
connect two identical Minkowski spacetimes possessing in general different masses and global charges.
 
Here we extend the solutions of Ref.~\cite{Dzhunushaliev:2025ntr} to the case of a system with a small throat parameter. 
As will be demonstrated,  this permits us to obtain configurations whose mass, observed at one end of the wormhole,
can be made of the order of the masses of the SM particles, despite the fact that the bare mass of the spinor field is of the order of  the Planck mass.
 The proposed mechanism for obtaining such realistic physical observable quantities like mass and charge is somewhat similar to the brane
scenario for solving the hierarchy problem. In both cases, there are two objects, on one of which the physical quantities have the Planck values, 
while on the other one these quantities (as a result of the action of  some mechanism) acquire the required values.
Within the brane scenario, these are two branes, while in our case these are two different asymptotically flat spacetimes connected by a wormhole. 
In particular, we will show that it is possible to obtain a localized one-particle solution 
whose physical characteristics (mass, charge, intrinsic angular momentum) correspond to the characteristics of an electron/positron.

Let us emphasise immediately that we are concerned here only with classical fields (spinor and electromagnetic). 
However,  a consideration of self-gravitating fermions 
still remains somewhat obscure, since a spinor field must be treated in terms of a normalizable quantum wave function. 
Nevertheless, one can impose certain restrictions by considering only one-particle fermion states 
and by ignoring second quantization of the fields. In this framework gravitational interaction can be treated purely classically. 
In this case the set of the Einstein-Dirac equations describes regular localized solitonic configurations~\cite{Finster:1998ws}, the so-called Dirac stars
\cite{Herdeiro:2019mbz,Dzhunushaliev:2018jhj,Dzhunushaliev:2019kiy,Blazquez-Salcedo:2019uqq,Herdeiro:2021jgc}, as well as 
 wormhole systems mentioned above~\cite{Konoplya:2021hsm}. 
Consistent with this, here we follow Ref.~\cite{Armendariz-Picon:2003wfx} where the conclusion has been drawn that a classical spinor field may 
appear either as a result of some effective description of a more complex quantum system or when a quantum state of a spinor is in some
sense ``close'' to a vacuum state where a classical consideration of a massive Dirac spinor may be a good
approximation. In doing so, it is 
assumed that the spinor field represents a set of four complex-valued spacetime functions which transform according to the
spinor representation of the Lorentz group~\cite{Armendariz-Picon:2003wfx}.  
So the mechanism described here can be thought of as some  (quasiclassical) approximation which could reflect the essence of a realistic quantum system.

In turn, it is known that classical spinors are Grassmann numbers, leading to difficulties in defining a current 
 $j^\mu$ and an energy-momentum tensor $T_{\mu}^\nu$. To overcome these difficulties, 
 in Appendix~\ref{app_1}, we show that it is possible to separate variables in spinors,
 taking Grassmann numbers as an individual factor. This enables one to introduce a mathematically well-defined Lagrangian 
 of a spinor field by integrating over Grassmann variables, analogous to what is done in supersymmetric theories.  
By varying such Lagrangian with respect to tetrads in a standard way, one can obtain a well-defined energy-momentum tensor
without Grassmann numbers. It can also be shown that a classical spinor field (whose components are Grassmann numbers)
appears as a result of the transition to the limit  $\hbar \rightarrow 0$ for the anticommutation relations of a quantum fermion field.

The paper is organized as follows.  In Sec.~\ref{prob_statem}, we present the action, {\it Ans\"{a}tze}, and field equations
for the configurations under consideration. In Sec.~\ref{quant_int}, we write down expressions for some physically interesting
quantities which will be useful for understanding the properties of the configurations in question. 
Sec.~\ref{num_sol_res} contains the results of numerical calculations, including an example of a configuration possessing the characteristics of an electron/positron.
Finally, in Sec.~\ref{concl}, we summarize the results obtained. 

\section{The model}
\label{prob_statem}

The total action for the system can be represented in the form [we use the metric signature $(+,-,-,-)$ and natural units $c=\hbar=1$]
\begin{equation}
\label{action_gen}
	S_{\text{tot}} = - \frac{1}{16\pi G}\int d^4 x
		\sqrt{- \cal g} R +S_{\text{sp}} +S_{\text{EM}},
\end{equation}
where $G$ is the Newtonian gravitational constant, $R$ is the scalar curvature, and $\cal g$ is the determinant of the metric; $S_{\text{sp}}$ and $S_{\text{EM}}$ are the actions of spinor, $\psi$,
and electromagnetic, $A_\mu$, fields, respectively. The action for the electromagnetic field can be derived from the Lagrangian 
$$
L_{\text{EM}}=-\frac{1}{4}F_{\mu\nu}F^{\mu\nu},
$$
where the electromagnetic field tensor is $F_{\mu\nu}=\partial_\mu A_\nu-\partial_\nu A_\mu$ and $\mu, \nu = 0, 1, 2, 3$ are spacetime indices.

In turn, the action  $S_{\text{sp}}$ for the spinor field $\psi$ appearing in Eq.~\eqref{action_gen} can be found from the Lagrangian 
\begin{equation}
	L_{\text{sp}} =	\frac{\imath}{2} \left(
			\bar \psi \gamma^\mu \psi_{; \mu} -
			\bar \psi_{; \mu} \gamma^\mu \psi
		\right) - m_s \bar \psi \psi ,
\label{lagr_sp}
\end{equation}
where $m_s$ is a bare mass of the spinor field and the semicolon denotes the covariant derivative defined as
$
\psi_{; \mu} =  [\partial_{ \mu} +1/8\, \omega_{a b \mu}\left( \gamma^a  \gamma^b- \gamma^b  \gamma^a\right) + \imath e A_\mu]\psi 
$ with $a,b$ being tetrad indices. 
Here $\gamma^a$ are the Dirac matrices in the Weyl representation in flat space, 
 $$
\gamma^0 =
     \begin{pmatrix}
        0   &   1 \\
        1   &   0
    \end{pmatrix},\quad
\gamma^k =
     \begin{pmatrix}
        0   &   \sigma^k \\
        -\sigma^k   &   0
    \end{pmatrix},
$$
where $k=1,2,3$ and $\sigma^k$ are the Pauli matrices.
  In turn, the Dirac matrices in curved space, $\gamma^\mu = e_a^{\phantom{a} \mu} \gamma^a$, 
 are obtained using the tetrad $ e_a^{\phantom{a} \mu}$, and $\omega_{a b \mu}$ is the spin connection [for its definition, see Ref.~\cite{Lawrie2002}, Eq.~(7.135)].
The gauge coupling constant $e$ describes the minimal interaction between the spinor
 and electromagnetic fields.

Then, by varying the action \eqref{action_gen} with respect to the metric, the spinor field, and the vector potential $A_\mu$, we obtain the Einstein, Dirac, and Maxwell field equations in curved spacetime:
\begin{eqnarray}
E_{\mu}^\nu	\equiv R_{\mu}^\nu - \frac{1}{2} \delta_{\mu }^\nu R - 8 \pi  G  \,T_{\mu }^\nu &=& 0	,
\label{feqs_10} \\
	\imath \gamma^\mu \psi_{;\mu} - \mu  \psi &=& 0 ,
\label{feqs_20}\\
	\imath \bar\psi_{;\mu} \gamma^\mu + \mu  \bar\psi &=&0 ,
\label{feqs_30}\\
\frac{1}{\sqrt{-\cal g}} \frac {\partial}{\partial x^\nu}
    \left(\sqrt{-\cal g}F^{\mu \nu}\right) &=& -e j^{\mu} ,
\label{feqs_40}
\end{eqnarray}
where $ j^{\mu}= \bar{\psi} \gamma^\mu \psi$ is the four-current of the spinor field.
The equation~\eqref{feqs_10} contains the energy-momentum tensor $T_{\mu}^\nu$, which can be represented in a symmetric form as
\begin{equation}
\label{EM_1}
	T_{\mu}^\nu = \frac{\imath }{4}g^{\nu\rho}
	\left[
		\bar\psi \gamma_{\mu} \psi_{;\rho} 
		+ \bar\psi\gamma_\rho\psi_{;\mu} - \bar\psi_{;\mu}\gamma_{\rho }\psi 
		- \bar\psi_{;\rho}\gamma_\mu\psi
	\right] - F^{\nu\rho} F_{\mu\rho}
    + \frac{1}{4} \delta_\mu^\nu F_{\alpha\beta} F^{\alpha\beta}.
\end{equation}

Since we consider here axially symmetric configurations,  we employ the following line element for a stationary, axially symmetric spacetime~\cite{Chew:2019lsa}:
\begin{equation}
\label{metric}
ds^2=e^f dt^2-e^{q-f}\left[e^b\left(dr^2+h d\theta^2\right)+h\sin^2\theta\left(d\varphi-\omega dt\right)^2\right],
\end{equation}
where the metric functions $f,q,b$, and $\omega$ depend solely on the radial coordinate $r$ and the polar angle $\theta$,  
and the auxiliary function $h = r^2 + r_0^2$ contains the throat parameter $r_0$; 
the radial coordinate $r$ covers the range $-\infty < r < + \infty$. 
The $z$-axis ($\theta=0$) represents the symmetry axis of the system. Asymptotically (as $r\to \pm \infty$), the functions $f, q, b,\omega \to 0$; i.e., 
the spacetime approaches Minkowski spacetime. 

The spinor field is parameterized by two complex functions
\cite{Herdeiro:2019mbz,Dzhunushaliev:2023vxy}
 \begin{equation}
    \psi^T = e^{\imath \left(M_\psi\varphi-\Omega t\right)}
        \begin{pmatrix}
            \psi_1, & \psi_2, & \psi_2^*, & \psi_1^*
        \end{pmatrix}\, .
\label{spinor}
\end{equation}
 Here the spinor frequency $\Omega$ is the eigenvalue of the Dirac Hamiltonian, $M_\psi$ is a half-integer parameter (the azimuthal number; in what follows we take $M_\psi=1/2$).
For our purposes, it is convenient to represent the components of the spinor field \eqref{spinor} as
$$
    \psi_1=\frac{1}{2}\left[X+Y+\imath\left(V+W\right)\right],\quad
    \psi_2=\frac{1}{2}\left[X-Y+\imath\left(V-W\right)\right],
$$
where the four real functions $X,Y,V$, and $W$ depend only on the coordinates $r$ and $\theta$.

The gauge field is parameterized by an electric, $\phi$, and a magnetic, $\sigma$, potentials
\begin{equation}
A_\mu=\{\phi(r,\theta),0,0,\sigma(r,\theta)\}.
\label{EM_ans}
\end{equation}
Hence, the electric field $\mathbf{E}$ and the magnetic field $\mathbf{H}$ as measured by the zero-angular-momentum observer (ZAMO) are given by
\begin{align}
\label{EM_components}
\begin{split}
&E_\beta=F_{\alpha\beta}n^\alpha=\left(0,-e^{-f/2}\left[\frac{\partial \phi}{\partial r}+\omega \frac{\partial \sigma}{\partial r}\right],
-e^{-f/2}\left[\frac{\partial \phi}{\partial \theta}+\omega \frac{\partial \sigma}{\partial \theta}\right],0\right) ,\\
&H_\beta=-\frac{1}{2}\epsilon_{\alpha\beta\mu\nu}n^\alpha F^{\mu\nu}=\left(0,\frac{1}{h}e^{(f-q)/2}\csc\theta \frac{\partial \sigma}{\partial \theta},
-e^{(f-q)/2}\csc\theta \frac{\partial \sigma}{\partial r},0\right) ,
\end{split}
\end{align}
where $n^\alpha$ is the four-velocity vector of the ZAMO, which in our case is
$$
n^\alpha=\sqrt{g^{tt}}\left(1,0,0,\frac{g^{t\varphi}}{g^{tt}}\right)=e^{-f/2}\left(1,0,0,\omega\right) .
$$

Then, substituting the {\it Ans\"{a}tze} \eqref{spinor} and \eqref{EM_ans}  and the metric  \eqref{metric} in the field equations \eqref{feqs_10}, \eqref{feqs_20}, and \eqref{feqs_40}, 
one can derive the corresponding set of ten field equations: four Einstein equations (elliptic partial differential equations) for the metric functions $f, q, b, \omega$, 
two Maxwell equations (also elliptic partial differential equations), and four Dirac equations. 
We do not show these cumbersome equations here  to avoid overburdening the text, but they can be found in Ref.~\cite{Dzhunushaliev:2025ntr}, 
where the procedure of solving this set of equations is described in detail as well.
For these equations, we impose the following boundary conditions
 for the metric functions at two spatial infinities ($x \to \pm\infty$) and
on the  $z$-axis ($\theta=0$ and $\theta=\pi$):
\begin{align}
\label{BCs_geom}
&\left. f \right|_{x \to \pm \infty} = \left. q \right|_{x \to \pm\infty} =\left. b \right|_{x \to \pm\infty} =
	\left. \omega \right|_{x \to\pm \infty} = 0  ; \\
&\left. \frac{\partial f}{\partial \theta}\right|_{\theta = 0,\pi} =
\left. \frac{\partial q}{\partial \theta}\right|_{\theta = 0,\pi} =
\left. b\right|_{\theta = 0,\pi} =
	\left. \frac{\partial \bar\omega}{\partial \theta}\right|_{\theta = 0,\pi} = 0. 
\end{align}
Note here that, in order to ensure the absence of a conical singularity, we must take $b|_{\theta=0,\pi}=0$ (the elementary flatness condition).
In turn, for the matter fields,  we take the boundary conditions
\begin{align}
\label{BCs_mat}
\begin{split}
&\left. \bar X \right|_{x \to\pm \infty} =
    \left. \bar Y \right|_{x \to\pm  \infty} =
  \left. \bar V \right|_{x \to\pm  \infty} =
  \left. \bar W \right|_{x \to\pm  \infty} =
    \left.  \bar\phi \right|_{x \to -  \infty} =
  \left.  \bar\sigma \right|_{x \to\pm  \infty} =   0,  \left.  \bar\phi \right|_{x \to +  \infty} =\bar{\phi}_{+\infty};
\\
&     \left. \frac{\partial \bar X}{\partial \theta}\right|_{\theta = 0} =
  \left. \frac{\partial \bar V}{\partial \theta}\right|_{\theta = 0} =
    \left. \frac{\partial  \bar\phi}{\partial \theta}\right|_{\theta = 0} =  0 ,  \left. \bar Y \right|_{\theta = 0} =\left. \bar W \right|_{\theta = 0}=\left.  \bar\sigma \right|_{\theta = 0}= 0 ;
\\
& \left. \frac{\partial \bar Y}{\partial \theta}\right|_{\theta = \pi} =
    \left. \frac{\partial \bar W}{\partial \theta}\right|_{\theta = \pi} =
    \left. \frac{\partial \bar \phi}{\partial \theta}\right|_{\theta = \pi} =  0 ,  \left.  \bar X \right|_{\theta = \pi} =\left.  \bar V \right|_{\theta = \pi}=\left.  \bar\sigma \right|_{\theta = \pi}= 0 ,
\end{split}
\end{align} 
where $\bar{\phi}_{+\infty}$ is an integration constant, see Eq.~\eqref{asympt_behav}.
Here and below we use the following dimensionless variables and parameters:
\begin{equation*}
\begin{split}
&x=m_s r, \quad
(\bar\Omega,\bar\omega)=(\Omega,\omega)/m_s, \quad
(\bar X, \bar Y, \bar V, \bar W)=\sqrt{\frac{4\pi G}{m_s}}(X,Y,V,W), \\
&\bar{\phi}=\sqrt{4\pi G}\phi, \quad
\bar{\sigma}=\sqrt{4\pi G} m_s\sigma, \quad
\bar{e}=\frac{e}{\sqrt{4\pi G} m_s} .
\label{dmls_var}
\end{split}
\end{equation*}

\section{Quantities of interest}
\label{quant_int}

Having obtained numerical solutions to the above set of ten field equations,
we can use them for calculating  physical characteristics of the configurations under consideration.

To calculate the angular momentum, we employ the expression for the angular momentum density coming from the  energy-momentum tensor~\eqref{EM_1}:
 \begin{align}
 \bar{T}_\varphi^t&\equiv \frac{4\pi G}{m_s}T_\varphi^t=\frac{1}{8 h}\Big\{4e^{-3f/2+q/2}h^{3/2}\sin\theta U_3\left[2\bar{\Omega}-2\bar{e}\bar{\phi}-\left(1+2\bar{e}\bar{\sigma}\right)\bar{\omega}\right]\nonumber\\
&+e^{-b/2-f/2}\sqrt{h}\Big\{2\sin\theta U_4\left[2x-h\left(2f_x-q_x\right)\right]
+\sqrt{h}\left[-2e^{b/2} U_1\left(1+2\bar{e}\bar{\sigma}\right)+U_2\left(2\cos\theta+\sin\theta\left[-2f_\theta+q_\theta\right]\right)\right]
\Big\}\nonumber\\
&+8 e^{-b-q}\left[\bar{\sigma}_\theta\bar{\phi}_\theta+\bar{\omega}\left(\bar{\sigma}^2_\theta+h\bar{\sigma}^2_x\right)+h\bar{\sigma}_x\bar{\phi}_x\right]
\Big\} ,
 \label{Ttphi_comp}
\end{align}
where 
$$
U_1=\bar{X}^2+\bar{Y}^2+\bar{V}^2+\bar{W}^2, \quad
U_2=\bar{X}^2-\bar{Y}^2+\bar{V}^2-\bar{W}^2, \quad
U_3=\bar{X}\bar{Y}+\bar{V}\bar{W}, \quad
U_4=\bar{X}\bar{W}-\bar{Y}\bar{V} .
$$
In Eq.~\eqref{Ttphi_comp}, the lower indices on the metric and field functions denote differentiation with respect to the corresponding coordinate.
Using \eqref{Ttphi_comp}, the total  dimensionless angular momentum reads
\begin{equation}
 \bar J\equiv \left(m_s/M_p\right)^2 J=-\frac{1}{2} \int_{-\infty}^{\infty} dx \int_{0}^{\pi} d\theta\, \bar{T}_\varphi^t\, e^{b-f+3 q/2} \left(x^2+x_0^2\right) \sin\theta  ,
\label{ang_mom_tot}
\end{equation}
where $M_p$ is the Planck mass.
The occurrence of a nonzero angular momentum \eqref{ang_mom_tot} is caused by the presence of a single spinor field carrying an intrinsic  half-integer momentum. 
In turn, the numerical calculations show that
the cumulative contribution coming from the electric and magnetic fields is equal to zero.  

The conserved total spinor charge associated with the  Noether current is derived using the expression for the spinor four-current $j^\mu$
[see after Eq.~\eqref{feqs_40}] in the form
\begin{equation}
Q_\psi=  \int  j^t \sqrt{{- \cal g}}\,dr d\theta d\varphi  = 
    \frac{1}{2}\left(\frac{M_p}{m_s}\right)^2 \int_{-\infty}^{\infty}dx\int_{0}^{\pi}d\theta\,
    e^{b+3(q-f)/2}
    \left(
        \bar X^2 + \bar Y^2 + \bar V^2 + \bar W^2
    \right) \left(x^2+x_0^2\right) \sin\theta  . 
\label{spinor_charge}
\end{equation}

Note that there exists a general relation for the angular momentum of a  spinor field \cite{Herdeiro:2019mbz},
\begin{equation}
	J = m Q_\psi .
\label{J_Q_rel}
\end{equation}
Our numerical computations confirm that for all solutions $m=1/2$, i.e., $m=M_\psi$, just as it should be.

Let us now consider the asymptotic behaviour of the solutions under investigation.
For the Maxwell fields we have
\begin{equation}
    \bar{\phi}\approx \bar{\phi}_{\pm \infty}+\frac{\bar{Q}_\pm}{x} +\cdots ,\quad
    \bar\sigma \approx - \frac{\bar{\mu}_{m\pm}}{x} \sin^2 \theta +\dots ,
\label{asympt_behav}
\end{equation}
where $\bar{\phi}_{\pm \infty}$ are two integration constants corresponding to the values of the electric field potential as $x\to \pm \infty$, respectively,
and $\bar Q_\pm\equiv  \sqrt{G/4\pi}m_s \,Q_\pm$ are the  charges of the configurations located to the left ($\bar Q_-$) and to the right ($\bar Q_+$) of the center
(for the physical interpretation of the constant $Q_-$, see Sec.~\ref{num_sol_res}).
Since the integration constants $\bar{\phi}_{\pm \infty}$ are arbitrary, here we put the constant  $\bar{\phi}_{-\infty}=0$, 
whereas the constant $\bar{\phi}_{+\infty}$ is so adjusted that 
the constraint equation~$\left(E^x_x-E^\theta_\theta\right)=0$ coming from the Einstein equations~\eqref{feqs_10}
is satisfied~\cite{Dzhunushaliev:2025ntr}. 
In turn, $\bar{\mu}_{m\pm}\equiv \sqrt{G/4\pi}m_s^2 \mu_{m\pm}$ is a dimensionless magnetic moment. 
Consequently,  the above quantities can be read off from the asymptotic expressions~\eqref{asympt_behav} as
\begin{equation}
    \bar{Q}_\pm = - \lim_{x\to\pm\infty} x^2 \frac{\partial\bar \phi}{\partial x}  , \quad
    \bar{\mu}_{m\pm} =\lim_{x\to\pm\infty} \frac{x^2}{\sin^2\theta} \frac{\partial\bar \sigma}{\partial x} . 
\label{asympt_charge}
\end{equation}

Further, asymptotic flatness of the spacetime implies that the metric approaches the Minkowski
metric at spatial infinity, i.e.,  $f,q,b, \omega\to 0$  asymptotically.
For the extraction of the global charges, it is necessary to consider the
behaviour of the metric functions at infinity,
\begin{equation}
f\approx \mp\frac{2\bar{M}_\pm}{x}+\cdots,\quad \bar\omega\approx \frac{2 \bar{J}_\pm}{x^3}+\cdots,
\label{asympt_metr}
\end{equation}
where  $\bar M_\pm\equiv m_s M_\pm/M_p^2$ and $\bar{J}_\pm\equiv \left(m_s/M_p\right)^2 J_\pm$ are dimensionless quantities. 
Then the ADM masses of the configuration $\bar{M}_\pm$ and the angular momenta $\bar{J}_\pm$ 
can be read off from the above expressions as
\begin{equation}
\label{expres_mass_mom}
\bar M_\pm=\pm\frac{1}{2}\lim_{x\to\pm\infty}x^2\partial_x f ,
\quad \bar{J}_\pm=\frac{1}{2}\lim_{x\to\pm\infty}x^3\bar{\omega} .
\end{equation}

Finally, the dimensionless gyromagnetic ratio $g$, expressed in units of $Q/\left(2 M\right)$, 
is defined from the relation
\begin{equation}
\mu_{m \pm}=g_\pm\frac{Q_\pm}{2}\frac{J_\pm}{M_\pm} 
\quad \Rightarrow \quad g_\pm=2\frac{\bar{\mu}_{m\pm} \bar{M}_{\pm}}{\bar{Q}_\pm \bar{J}_{\pm}}.
\label{gyro}
\end{equation}

\section{Numerical calculations}

\subsection{Numerical method}
\label{num_meth}

We solve the corresponding set of mixed order Einstein-Dirac-Maxwell PDEs (see Ref.~\cite{Dzhunushaliev:2025ntr}) 
with the boundary conditions~\eqref{BCs_geom}-\eqref{BCs_mat}. 
In order to map the infinite range of the radial variable $x$ 
to the finite interval, we introduce the compactified coordinate~$\bar x$ as 
\begin{equation}
     x = c_k\frac{\bar x}{\left(1-\bar x^2\right)^2} \, ,
\label{comp_coord}
\end{equation}
which maps the infinite region $(-\infty;\infty)$ onto the finite interval $[-1; 1]$. Here $c_k$ is a constant which is used to adjust the contraction of
the grid. In our calculations, we typically take $c_k=1$.

Technically, the PDEs  are discretized on some grid,
and the resulting set of nonlinear algebraic equations 
is then solved by using a modified Newton method.
The underlying linear system is solved 
with the Intel MKL PARDISO sparse direct solver~\cite{pardiso} 
and the CESDSOL library\footnote{Complex Equations-Simple Domain 
partial differential equations SOLver, a C++ package developed by I.~Perapechka,
see Refs.~\cite{Herdeiro:2019mbz,Herdeiro:2021jgc}.}.
The package provides an iterative procedure to obtain an exact solution starting from some initial guess configuration. 
As such a configuration, we take a nonrotating system found in Ref.~\cite{Dzhunushaliev:2025lki}.
In doing so, typical mesh sizes include $400\times 400$  points covering  the integration region $-1\leq \bar x \leq 1$ 
[given by the compact radial coordinate~\eqref{comp_coord}] and $0\leq \theta \leq \pi$. 
The resulting system is solved iteratively until convergence is achieved.
In all cases, the typical errors are of order of $10^{-4}$.
In turn, the accuracy of numerical solutions is also verified by comparing them with the asymptotic expressions~\eqref{asympt_behav} and \eqref{asympt_metr}.

\subsection{Numerical results}
\label{num_sol_res}

The search for finite-energy solutions to the set of the Einstein, Dirac and Maxwell equations  is carried out  by assigning different 
values of the free system parameters  $x_0$,  $\bar\Omega$, and $\bar{e}$ for which one can find regular solutions. 
For our purposes, in the present paper we restrict ourselves only to a case with uncoupled spinor field ($\bar{e}=0$) and 
consider solutions in a restricted range of spinor frequencies $-1 \lesssim \bar{\Omega} \leq -0.9$ (a more complete set of solutions can be found in Ref.~\cite{Dzhunushaliev:2025ntr}). 
For these values of the  input parameters, we construct solutions for small values of
the throat parameter $0.025 \leq x_0 \leq 0.2$. Note that when $\bar{\Omega} \to -1$ the accuracy of numerical calculations decreases;
therefore, in the figures below we show only solutions that were obtained with the required numerical accuracy, see Ref.~\cite{Dzhunushaliev:2025ntr}.

\begin{figure}[t]
    \begin{center}
        \includegraphics[width=.9\linewidth]{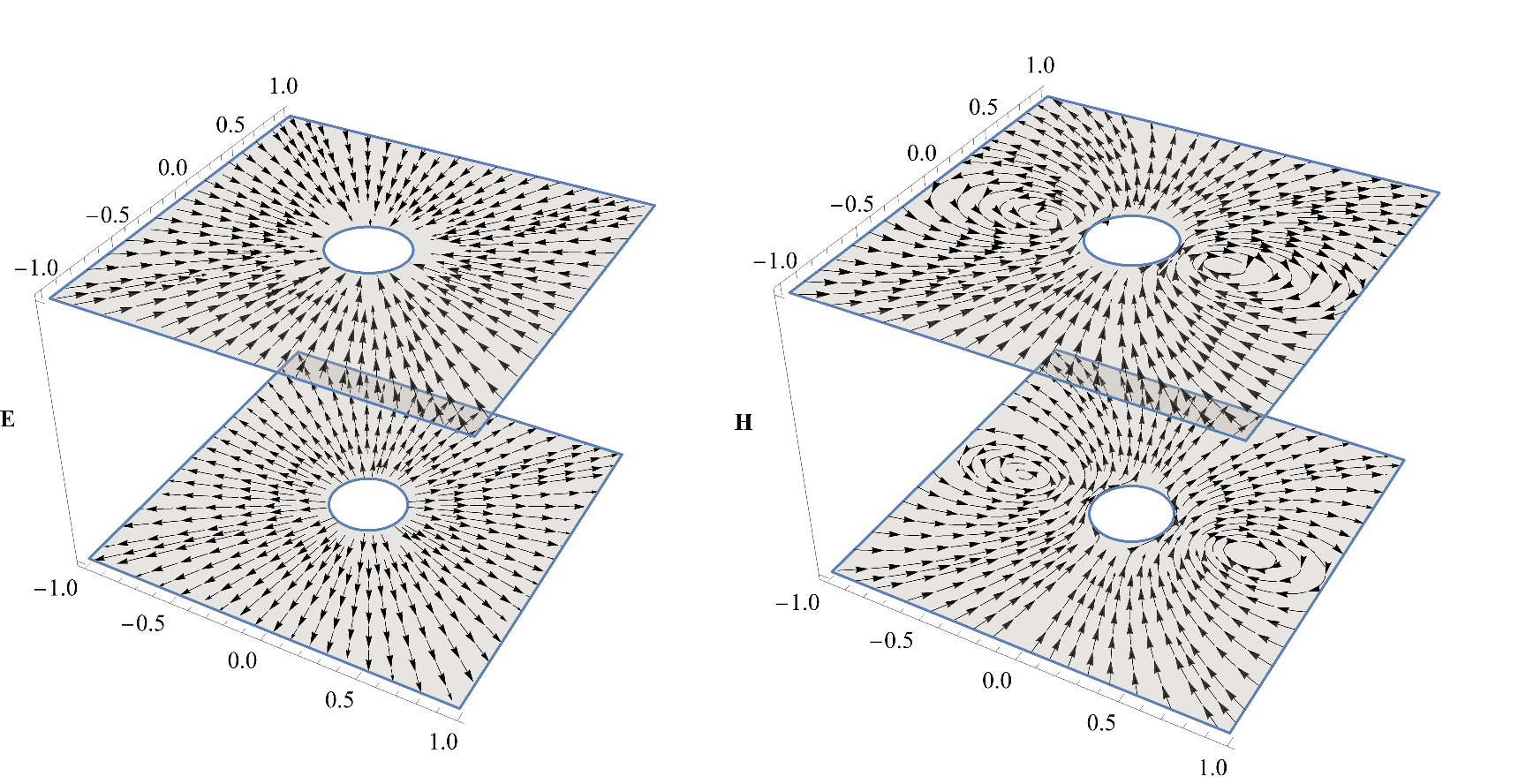}
    \end{center}
    \vspace{-.5cm}
    \caption{The typical distributions of
    lines of force of the dimensionless electric, $\mathbf{\bar E}\equiv\sqrt{4\pi G}/m_s \mathbf{E}$, and magnetic, $\mathbf{\bar H}\equiv\sqrt{4\pi G}/m_s \mathbf{H}$, fields~\cite{Dzhunushaliev:2025ntr}. 
    The top plots correspond to the subspace with $x>0$, while the bottom plots are for the subspace with $x<0$. }
\label{fig_EMfields}
\end{figure}

\begin{figure}[t]
    \begin{center}
        \includegraphics[width=.49\linewidth]{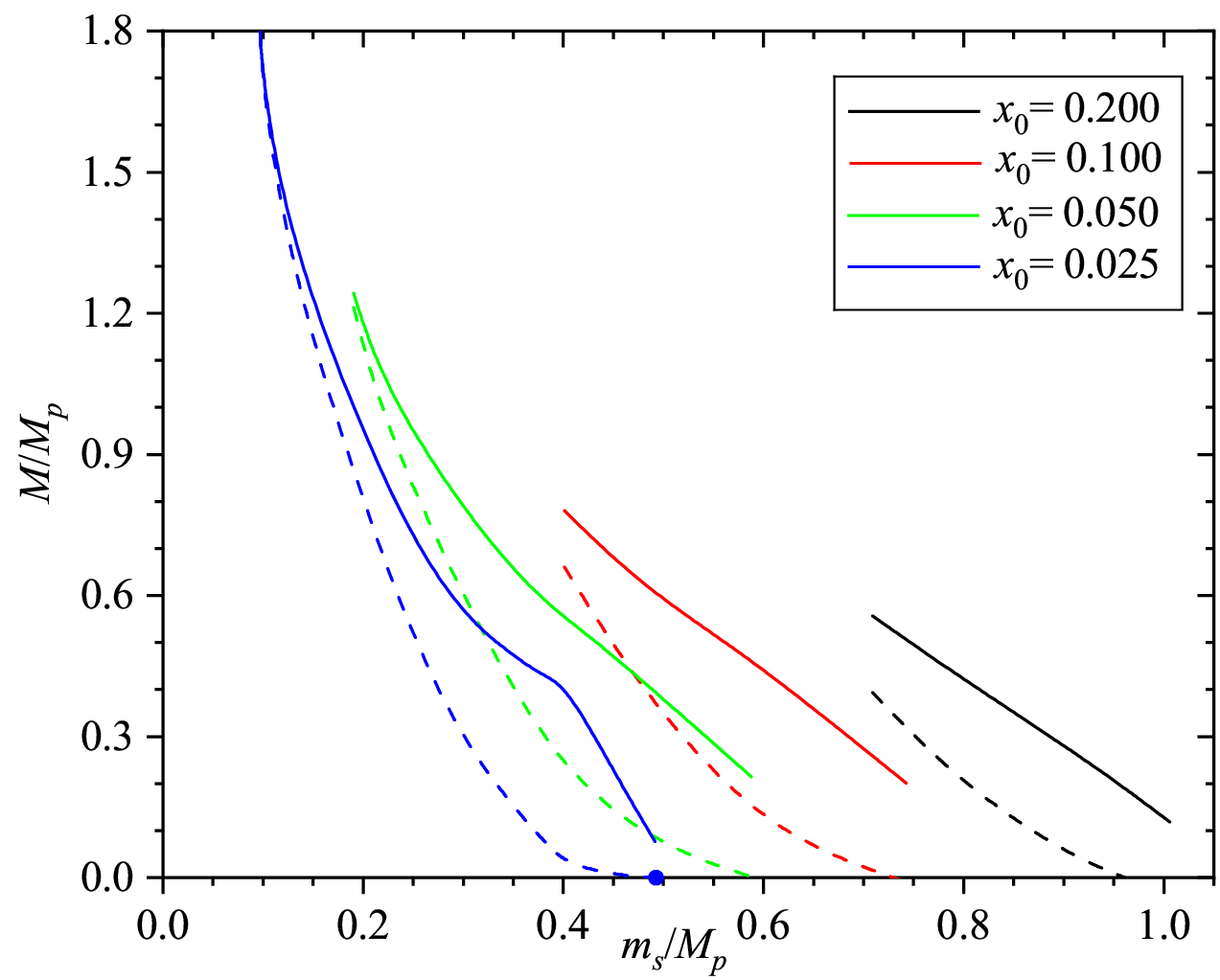}
        \includegraphics[width=.49\linewidth]{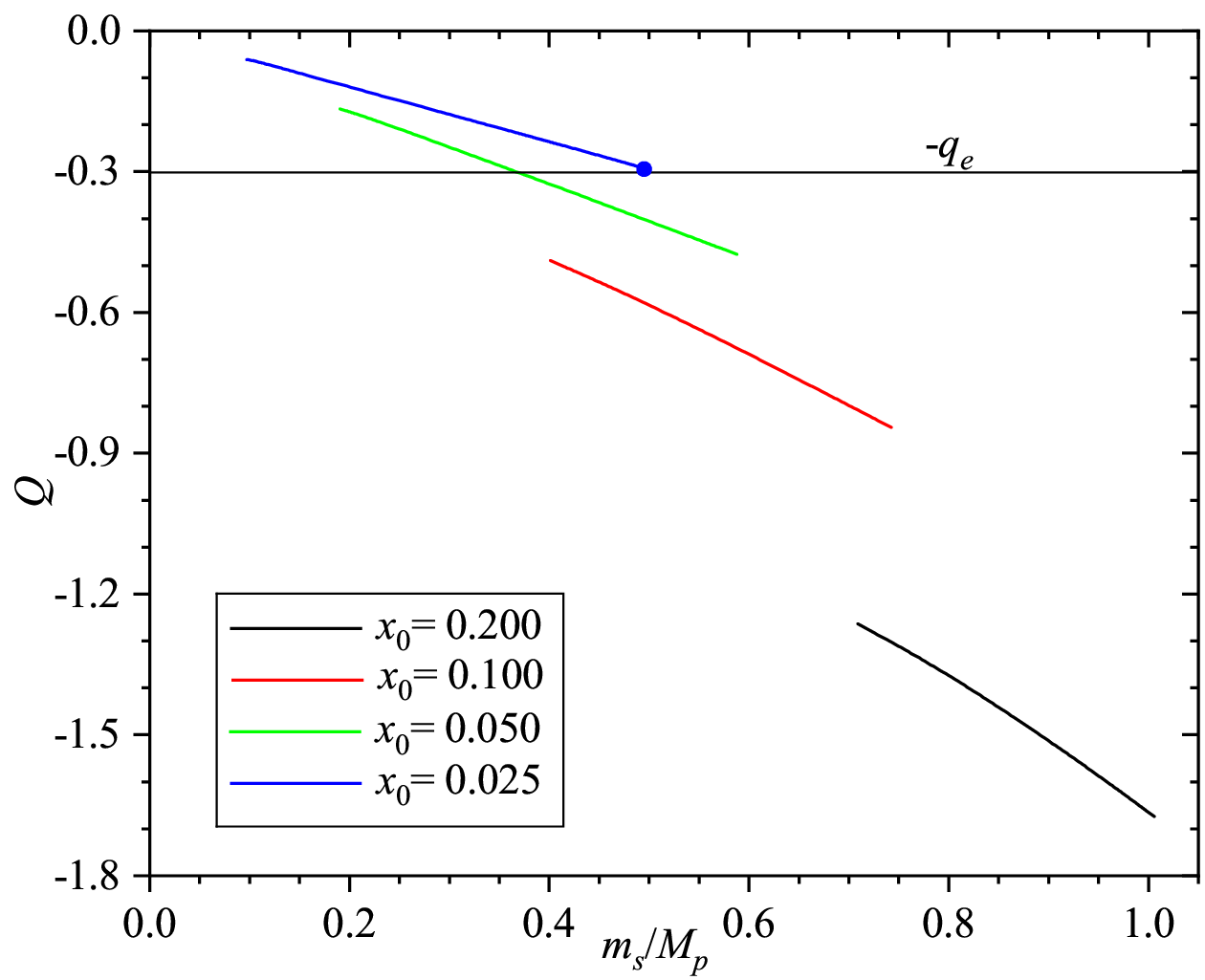}
    \end{center}
    \vspace{-.5cm}
    \caption{ The masses $M_+$  and $M_-$  (the left panel) and the electric charges $Q=Q_+=-Q_-$ (the right panel) vs. the mass of the spinor field $m_s$ are shown for different values 
    of the throat parameter $x_0$ and with $\bar{e}=0$.
The solid lines correspond to the systems in the subspace with $x>0$, while the dashed lines are for the configurations in the subspace with $x<0$. 
The leftmost points of the curves correspond to the systems with $\bar{\Omega}=-0.9$, while the rightmost points are for the systems with $\bar{\Omega}\to -1$ (cf. Fig.~\ref{fig_J_g}).
The bold dots label the configuration with the mass and charge of a positron.}
    \label{fig_Mass_Charge}
\end{figure}

\begin{figure}[t]
    \begin{center}
        \includegraphics[width=.49\linewidth]{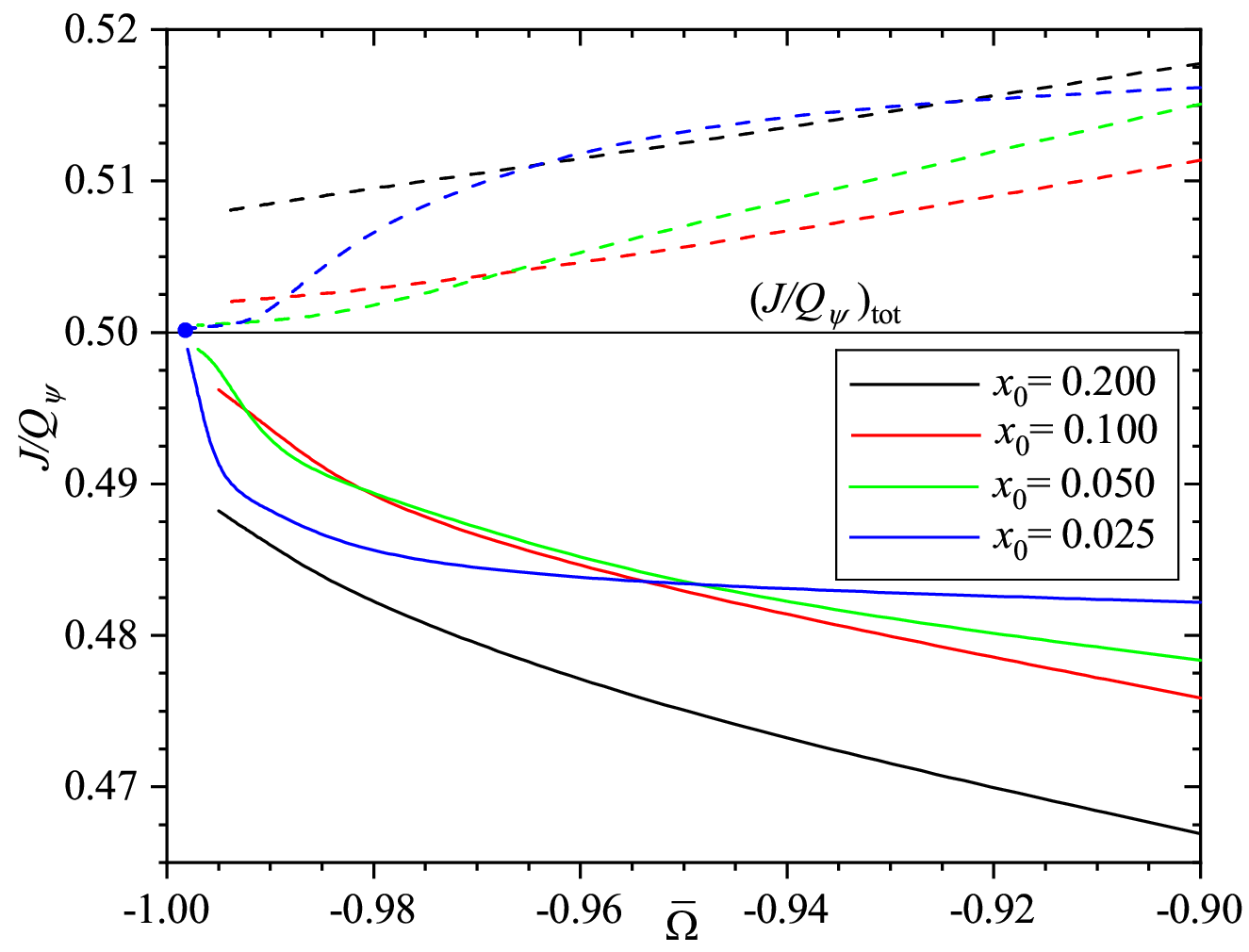}
        \includegraphics[width=.49\linewidth]{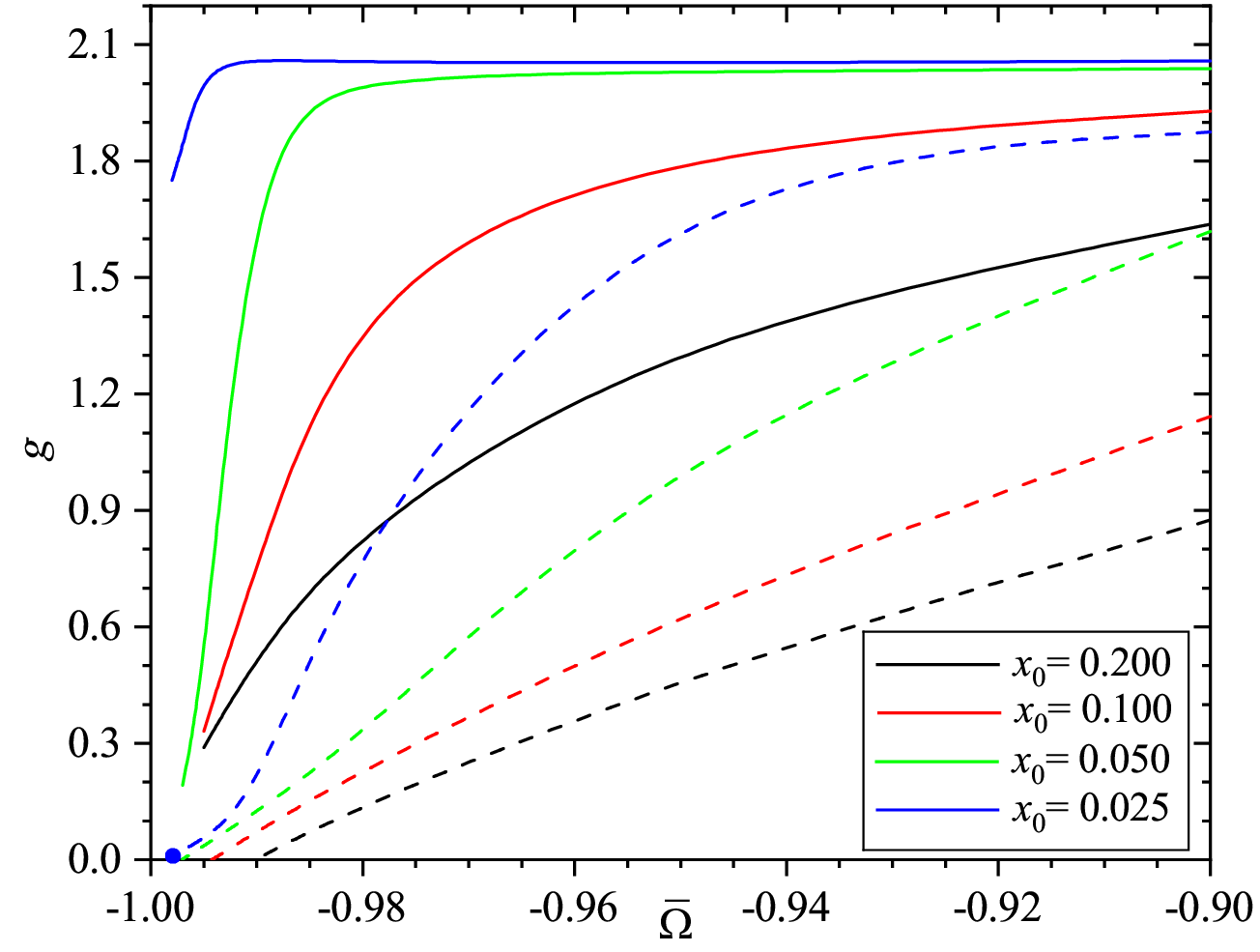}
    \end{center}
    \vspace{-.5cm}
    \caption{ The ratio of the angular momentum $J$ to the Noether charge $Q_\psi$ (the left panel) and the gyromagnetic ratio  $g$ (the right panel) vs. the spinor frequency $\bar{\Omega}$  are shown for different values 
    of the throat parameter $x_0$ and with $\bar{e}=0$.
The solid lines correspond to the systems in the subspace with $x>0$, while the dashed lines are for the configurations in the subspace with $x<0$. 
The bold dots label the configuration with the mass and charge of a positron.}
    \label{fig_J_g}
\end{figure}

Notice that since here we construct solutions for classical fields, the Noether charge $Q_\psi$ calculated using~\eqref{spinor_charge} 
is some arbitrary number whose value in general is determined by the specific values of the input parameters  $x_0$,  $\bar\Omega$, and $\bar{e}$,
as well as by the mass of the spinor field $m_s$. 
However, if one goes beyond the classical treatment of the fields
and imposes the quantum nature of fermions, this requires that $Q_\psi=1$~-- the one-particle condition. 
Since in the present paper we are interested precisely in one-particle solutions (see Introduction), 
all results given below correspond to such solutions, i.e., we always impose 
the normalization condition $Q_\psi=1$. This results in the fact that each of sequences of solutions 
for a fixed $x_0$ shown in the figures below corresponds to configurations with constant $Q_\psi=1$ and varying mass of the spinor field $m_s$ (that is, the curves
correspond to sequences of solutions of different models).

Before discussing the results of numerical calculations, we also note that 
for the physical interpretation of the constant $Q_-$ appearing in Eq.~\eqref{asympt_behav} it is necessary to consider the behaviour of electric lines of force calculated using Eq.~\eqref{EM_components}
and shown in Fig.~\ref{fig_EMfields}.
These plots are made in a meridional plane $\varphi=\text{const.}$
    spanned by the coordinates $u=\sqrt{x^2+x_0^2}\sin\theta$ and $v=\sqrt{x^2+x_0^2}\cos\theta$. Such coordinates imply that the whole
    space is divided into two subspaces located at $x>0$ and $x<0$, and they are joined together  along a circle with radius $x_0$~\cite{Dzhunushaliev:2025ntr}.
It is seen from this figure that for a distant observer located in the subspace with $x < 0$ the center of the wormhole looks like a positive charge.
In this connection, for the correct physical interpretation of $\bar Q_-$, the expression $\bar Q_-/x$ in Eq.~\eqref{asympt_behav} must be rewritten in the form $-\bar Q_-/(-x)$,
in order for the denominator $(-x)$ to be a positive number.  This means that a distant observer in the region with $x<0$ will observe an electric charge equal to $-\bar Q_-$.

Keeping all this in mind, the corresponding results of numerical calculations are shown in Figs.~\ref{fig_Mass_Charge} and~\ref {fig_J_g}. Fig.~\ref{fig_Mass_Charge}
shows the dependencies of the masses $M_+$ and $M_-$ [calculated using the asymptotic expression given by Eq.~\eqref{expres_mass_mom}] 
and of the electric charges $Q=Q_+=-Q_-$ [calculated using the asymptotic expression given by Eq.~\eqref{asympt_charge}] 
on the mass of the spinor field $m_s$ expressed in Planck units. It is seen from these graphs that the masses of the configurations located in the subspace with $x>0$
are always positive and are of the order of $M_p$. In turn, the masses of the systems in the subspace with $x<0$ tend to zero at some value of the mass of the spinor field $m_s\sim M_p$
and, as the numerical calculations indicate, they can even be negative  (cf. the case of nonrotating systems considered in Ref.~\cite{Dzhunushaliev:2025lki}).
\emph{This enables one to obtain configurations whose observed mass can be made equal to a mass typical of that of the SM particles, whereas the bare mass 
of the spinor field  $m_s$ appearing in the Lagrangian~\eqref{lagr_sp} continues to be of the order of the fundamental Planck mass. }

The next step is to find the required value of an electric charge of the modeled system. If, for example, we wish to obtain an elementary charge  $q_e\approx 0.3$ (in natural units),
this can be done by suitably adjusting a value of one of free system parameters~-- the throat parameter $x_0$. The right panel of Fig.~\ref{fig_Mass_Charge} shows
the change in magnitude of the electric charge $Q$ for different values of $x_0$. It is seen from this figure that the required charge
 $Q\approx -q_e$ can be obtained for $x_0=0.025$. The configuration whose mass and charge correspond to the parameters of a positron is labeled in the left and right panels 
of Fig.~\ref{fig_Mass_Charge} by the bold dot located at $m_s/M_p\approx 0.49$.

Consider now the behaviour of the angular momentum. The numerical calculations indicate that for the systems under consideration 
the ratio of the total angular momentum 
$J$ [calculated using  Eq.~\eqref{ang_mom_tot}] to the total Noether charge [calculated using  Eq.~\eqref{spinor_charge}] is $\left(J/Q_\psi\right)_{\text{tot}}=1/2$, see Eq.~\eqref{J_Q_rel}.
In turn, the magnitude of this ratio as measured by distant observers located at spatial infinities, i.e., $J_\pm/Q_{\psi \pm}$, 
can be found from Eq.~\eqref{spinor_charge} and the asymptotic expression for $J_\pm$ given by  Eq.~\eqref{expres_mass_mom}. In doing so, 
the values of $Q_{\psi +}$ and $Q_{\psi -}$ are obtained by integrating in~\eqref{spinor_charge} in the subspaces with $x>0$ and $x<0$, respectively,
and, in performing the integration, the lower/upper limit of the integral in the subspace with $x>0$/$x<0$ corresponds to the point $x=x_{\text{max}}\neq 0$ 
where the maximum of the charge density of the spinor field $j^t$ appearing in Eq.~\eqref{spinor_charge} is located~\cite{Dzhunushaliev:2025lki}.
Strictly speaking, for the axially symmetric configurations under consideration, $x_{\text{max}}$ is a function of the angular variable $\theta$, 
but for the values of the system parameters considered here (i.e., $0.025 \leq x_0 \leq 0.2$ and $-1 \lesssim \bar{\Omega} \leq -0.9$) the value of $x_{\text{max}}$ is practically constant
and approximately equal to zero, $x_{\text{max}}\to -0$. 
Consistent with this, the corresponding results of calculations are shown in the left panel of Fig.~\ref{fig_J_g}.
One can see that in the subspace with $x>0$ the ratio $J_+/Q_{\psi +}$ is always less than 1/2, while in the subspace with $x<0$ the ratio $J_-/Q_{\psi -}$ always exceeds 1/2.
For the system whose parameters correspond to the characteristics of a~positron, the ratio $J_-/Q_{\psi -}\approx 1/2$ (labeled by the bold dot).

Finally, consider the behaviour of the gyromagnetic ratio  $g$ given by Eq.~\eqref{gyro} for the configurations under investigation for different values of $x_0$, see the right panel of Fig.~\ref{fig_J_g}.
It is seen from this figure that the value of $g$ depends considerably both on the value of the throat parameter $x_0$ and on the spinor frequency $\bar{\Omega}$:
as $x_0$ decreases, the magnitude of  $g$ increases  (for a fixed $\bar{\Omega}$). However, for the system possessing parameters of a positron (labeled by the bold dot in the right panel of Fig.~\ref{fig_J_g}) 
 the gyromagnetic ratio $g\to 0$, whereas for the electron/positron $g\approx 2$.

\section{Conclusion}
\label{concl}

In the present paper we have considered a possible mechanism for obtaining the values of mass of localized field configurations of the order  of the masses
of the SM particles, starting from the parent Lagrangian in which a bare mass of a fermion has the value of the order of the fundamental Planck mass.
This may be achieved because the system has a nontrivial wormholelike spacetime topology. Such a topology is supported by a non-phantom spinor field, 
which also provides the presence of nonzero angular momentum. In turn, the presence in the system of an electric field  possessing an asymptotically Coulomb behaviour
provides the presence of an electric charge.  

Let us enumerate the main results obtained:
\begin{itemize}
\item[(i)] We have introduced a strict, mathematically correct definition  of the Lagrangian for a classical spinor field 
and shown that in the limit  $\hbar \rightarrow 0$ an anticommuting quantum spinor field is transformed into a classical spinor field (for details, see Appendix~\ref{app_1}).

\item[(ii)] We have found regular asymptotically flat solutions describing gravitating configurations with a wormhole topology. 
The configurations are asymmetric with respect to the center of the system $x=0$; this in general results in the fact that they 
have different global charges (masses, electric charges, angular momenta) at two asymptotically flat ends. Correspondingly, the whole space 
can be divided into two subspaces possessing different physical characteristics. 
Such configurations can be thought of as wormholes connecting two different universes, and in our universe (to which there corresponds the subspace with $x<0$ in the case considered here)
the required masses of the order of those of the SM particles are observed.

\item[(iii)] The physical properties of the resulting configurations are completely determined by the values of three input quantities: the throat parameter, $x_0$, the spinor frequency, $\bar\Omega$, and
the electromagnetic coupling constant, $\bar e$. We have shown that for the uncoupled spinor field (i.e. when $\bar e=0$), by adjusting the values of $\bar\Omega$ and $x_0$,
one can provide the required mass of the system of the order of the masses of the SM particles and the electric charge equal, for example, to the charge of a positron.
These mass and charge are observed at one end of the wormhole, whereas at the other end of the wormhole these physical quantities acquire the Planck values.

\item[(iv)] Due to the presence of the spinor field, the total angular momentum of the system is equal to 1/2. Although 
 the angular momenta as measured by distant observes (located at spatial infinities)
are not strictly equal to 1/2, for the configuration with the mass and charge of a positron considered here 
the intrinsic angular momentum is approximately equal to 1/2. 

\item[(v)] The gyromagnetic ratio $g$ differs considerably from the value $g\approx 2$ that is inherent to an electron/positron.
\end{itemize}

Thus we have obtained a wormholelike configuration, at one asymptotic end of which a distant observer  (located at  $x < 0$) sees a particlelike object
whose mass, charge, and intrinsic angular momentum are comparable to those of a positron. For such a system 
the electric field enters the wormhole from one asymptotically flat spacetime with the radial coordinate $x > 0$ and exits to another asymptotically flat spacetime 
with $x < 0$~\cite{Dzhunushaliev:2025ntr}. This configuration is obtained as a solution of the source-free Maxwell equations ($e=0$), coupled to Einstein gravity, with the seasoning of nontrivial topology. 
This means that the solution obtained incarnates Wheeler's idea of ``mass without mass'' and 
``charge without charge''~\cite{Wheeler}, and can be thought of as a model of \emph{a classical charge possessing a spin}.

\section*{Acknowledgements}

We gratefully acknowledge support provided by the program
No.~BR24992891 (Integrated research in nuclear, radiation physics and engineering, high energy physics and cosmology for the development of competitive technologies)
of the Committee of Science of the Ministry of Science and Higher Education of the Republic of Kazakhstan.

\appendix

\section{Lagrangian for classical spinors with Grassmann numbers}
\label{app_1}

To begin with, let us recall the definition of a spinor that can be found in standard textbooks, see, e.g., Ref.~\cite{Shifman}. A Weyl spinor is a two complex component object 
$\chi = \begin{pmatrix}
	\chi_1 \\
	\chi_2 
\end{pmatrix} \in \mathbf{C}^2$ transforming under an element 
$
	\mathcal{M} \in \mathrm{Sl}(2, \mathbf{C})
$ as 
$$
	\chi_\alpha \rightarrow \chi^\prime_\alpha = \mathcal{M}_\alpha^{\phantom{\alpha} \beta} \chi_\beta , 
$$
with $\alpha, \beta$ labeling the components. A two-component object $\bar{\chi}$ belonging to the complex conjugate representation 
 and transforming as
$$
	\bar{\chi}_{\dot{\alpha}} \rightarrow \bar{\chi}^\prime_{\dot{\alpha}} = 
	\left(\mathcal{M}^*\right)_{\dot{\alpha}}^{\phantom{\alpha} \dot{\beta}} 
	\bar{\chi}_{\dot{\beta}}
$$
is called a dotted spinor.

The scalar products $\chi \eta$ and $\bar{\chi} \bar{\eta}$ are defined as
\begin{equation}
	\chi \eta := \chi^{\alpha} \eta_{\alpha} = \eta \chi, \quad 
	\bar{\chi} \bar{\eta} := \bar{\chi}_{\dot{\alpha}} \bar{\eta}^{\dot{\alpha}} 
	= \bar{\eta} \bar{\chi} . 
\label{scalar_prod}
\end{equation}
For the scalar product \eqref{scalar_prod} to be symmetric, we find 
\begin{equation}
	\chi_\alpha \eta_\beta = - \eta_\beta \chi_\alpha , 
\label{Grassman_spinor}
\end{equation}
i.e., spinors must anticommute; this means that the components of the Weyl spinor are Grassmann numbers.

A Dirac spinor $\Psi$ is defined to be the direct sum of two Weyl spinors $\chi$ and $\bar{\eta}$ of opposite chirality 
\begin{equation}
	\Psi := 
	 \begin{pmatrix}
		\chi_\alpha \\
		\bar{\eta}^{\dot{\alpha}}  
	\end{pmatrix}, \quad 
	\bar{\Psi} =  
	 \begin{pmatrix}
		\eta^\alpha , \bar{\chi}_{\dot{\alpha}}  
	\end{pmatrix} . 
\label{Dirac_spinor}
\end{equation}

For a correct definition of the Lagrangian of a spinor field, let us consider the expression
\begin{equation}
	L_D = \imath \bar{\Psi} \gamma^{\mu} D_\mu \Psi - m \bar{\Psi} \Psi 
	=\imath \bar{\Psi} 	 \begin{pmatrix}
			0																&	\sigma^\mu \\
			\bar{\sigma}^\mu	& 0
		\end{pmatrix} D_\mu \Psi - m \bar{\Psi} \Psi 
	= \imath \left[ 
		\bar{\chi}_{\dot{\beta}} \left( \bar{\sigma}\right) ^{\dot{\beta} \alpha} D_\mu \chi_{\alpha} 
		+ \eta^{\alpha} \left( \sigma^\mu \right)_{\alpha \dot{\beta}} D_\mu {\bar{\eta}}^{\dot{\beta}} 
	\right] 
	- m \left( \eta^{\alpha} \chi_{\alpha} + \bar{\chi}_{\dot{\alpha}} \bar{\eta}^{\dot{\alpha}}\right) , 
\label{Dirac_Lagr}
\end{equation}
where 
$\sigma^\mu := \left(\mathbb{I}, \sigma^k\right)$ and $\bar\sigma^\mu := \left(\mathbb{I}, -\sigma^k\right)$, 
and $\chi_\alpha$ and $\bar{\eta}^{\dot{\alpha}}$ are Grassmann numbers.

In Refs.~\cite{Shifman} (Sec. 7.1) and~\cite{Raja} (Sec. 9.1), 
it is suggested that the separation of variables is carried out such that one can explicitly extract Grassmann numbers,
$$
	\Psi_\alpha = \sum_{i} \tilde{\Psi}_{i} \theta_{i \alpha} ,
$$
where $\alpha$ is the spinorial index and the coeﬃcients $\theta_{i \alpha}$ are Grassmann numbers.

For our purposes and in order to take into account the fact that spinors \eqref{Dirac_spinor} are Grassmann numbers \eqref{Grassman_spinor}, 
we choose another decomposition of spinor components into ordinary functions and Grassmann numbers in the following form:
$$
	\Psi := 
	 \begin{pmatrix}
		\chi_\alpha \\
		\bar{\eta}^{\dot{\alpha}}  
	\end{pmatrix} \theta, \quad 
	\bar{\Psi} =  
	 \begin{pmatrix}
		\eta^\alpha , \bar{\chi}_{\dot{\alpha}}  
	\end{pmatrix} \theta^*  ,
$$
where $\theta$ and $\theta^*$ are independent Grassmann numbers, while $\chi_\alpha$ and $\bar{\eta}^{\dot{\alpha}}$ are now complex functions.
For such a separation of variables into complex and  Grassmann numbers, the Lagrangian \eqref{Dirac_Lagr} takes the form
$$
	\tilde{L}_D = \left\lbrace \imath 
	\left[ 
		\bar{\chi}_{\dot{\beta}} \left( \bar{\sigma}\right) ^{\dot{\beta} \alpha} D_\mu \chi_{\alpha} 
		+ \eta^{\alpha} \left( \sigma^\mu \right)_{\alpha \dot{\beta}} D_\mu {\bar{\eta}}^{\dot{\beta}} 
	\right] 
	- m \left( \eta^{\alpha} \chi_{\alpha} + \bar{\chi}_{\dot{\alpha}} \bar{\eta}^{\dot{\alpha}}\right) 
	\right\rbrace 
	\theta^* \theta .
$$
Then, introducing integration over the Grassmann variables according to the general rules
$$
	\int 1 d \theta = 0, \quad \int \theta d \theta = 1, \quad 
	\int d \theta d \theta^* \left( \theta^* \theta \right) = 1 , 
$$
one can write out the Lagrangian for a classical spinor field  using integration over the Grassmann variables:
\begin{equation}
	\mathcal{L}_D = \int \tilde{L}_D d \theta d \theta^* 
	= \imath \left[ \bar{\chi}_{\dot{\beta}} \left( \bar{\sigma}\right) ^{\dot{\beta} \alpha} D_\mu \chi_{\alpha} 
		+ \eta^{\alpha} \left( \sigma^\mu \right)_{\alpha \dot{\beta}} D_\mu {\bar{\eta}}^{\dot{\beta}} 
	\right] 
	- m \left( \eta^{\alpha} \chi_{\alpha} + \bar{\chi}_{\dot{\alpha}} \bar{\eta}^{\dot{\alpha}}\right) 
	+ (h. c.) .
\label{Lagrangain_spinor_classic}
\end{equation}
\emph{This gives us a strict, mathematically correct definition of the Lagrangian for a classical spinor field, enabling one 
to define correctly the current and the energy-momentum tensor for the classical spinor field in terms of ordinary $c$-numbers. }
Notice that the integration over the Grassmann variables in \eqref{Lagrangain_spinor_classic} is significantly simpler 
than that in a path integral when fermion fields are present (see, e.g., Ref.~\cite{Raja}).
In this connection also note that, as pointed out in Ref.~\cite{Raja}: ``In the study of the non-second-quantised Dirac equation, 
standard textbooks usually treat the `Dirac wavefunction' as a $c$-number spinor field, without bothering about Grassmann fields.''

Now it is necessary to address the question in what sense one should understand the limiting 
transition $\hbar \rightarrow 0 $ from a quantum spinor field to a classical one. As we know, 
the anticommutation relations for operators of a fermion field are of the form
$$
	\left\lbrace \hat{\Psi}_\alpha \left( x^\mu\right) , \hat{\Psi}_\beta \left(y^\mu\right) 
	\right\rbrace = \frac{1}{\imath} S_{\alpha \beta} \left(x^\mu - y^\mu\right) 
	= \left( 
		\imath \slashed{\nabla}  - m
	\right)_{\alpha \beta} 
	\int \frac{d^4 k}{\left(2 \pi \right)^3 } \epsilon\left( k_0\right) 
	\delta \left(k^2 - m^2 \right) 
	e^{- \imath k \cdot \left(x - y\right) } ,
$$
where $S_{\alpha \beta}$ is a fermion propagator and $\epsilon\left( k_0\right) $ is a step function ($\epsilon=1$ for $k_0>0$ and $\epsilon=-1$ for $k_0<0$).
This expression is written in natural units; if we restore the constants $\hbar$ and $c$,  Planck's constant will appear in the Dirac 
delta function as
$\delta \left(\hbar^2 k^2 - m^2 c^2\right) $. When $\hbar \rightarrow 0$, we have
$
	\delta \left(\hbar^2 k^2 - m^2 c^2\right) \rightarrow \delta \left(- m^2 c^2\right) = 0 
$. This means that in the limit  $\hbar \rightarrow 0$ the anticommutator 
$
\left\lbrace \hat{\Psi}_\alpha \left( x^\mu\right) , \hat{\Psi}_\beta \left(y^\mu\right) \right\rbrace = 0
$, and this in turn means that the components of the spinor $\Psi$ become Grassmann numbers.

Thus, let us emphasise the following results:
\begin{itemize}
\item It is shown that in order to take into account the fact that the components of a spinor are Grassmann numbers,
one can represent the spinor as the product of ordinary functions and Grassmann numbers. 
\item We gave a strict, mathematically correct definition of the Lagrangian of a spinor field arising after integration over Grassmann variables.
Such approach enables one to define correctly the current and the energy-momentum tensor for the classical spinor field.
\item It is shown that in the limit $\hbar \rightarrow 0$ an anticommuting quantum spinor field is transformed into a classical spinor field whose components are Grassmann numbers.
\end{itemize}

\end{document}